\documentclass[a4paper]{article}
\usepackage[utf8]{inputenc}
\usepackage[T1]{fontenc}
\usepackage{multirow}
\usepackage{tabularx}
\usepackage{float}
\usepackage{xcolor}

\usepackage{INTERSPEECH2020}

\usepackage[square,numbers]{natbib} 
\usepackage[italic]{mathastext} 
\usepackage[hidelinks]{hyperref} 
\usepackage{microtype}  
\usepackage{enumitem} 

\usepackage{amsthm}
\theoremstyle{plain}

\makeatletter
\renewcommand{\paragraph}{%
  \@startsection{paragraph}{4}%
  {\z@}{0.6ex \@plus 1ex \@minus 0.2ex}{-1em}%
  {\normalfont\normalsize\bfseries}%
}
\makeatother

\title{SpeedySpeech: Efficient Neural Speech Synthesis}
\name{Jan Vainer, Ondřej Dušek}
\address{Charles University, Faculty of Mathematics and Physics, Prague, Czechia}
\email{vainerjan@gmail.com, odusek@ufal.mff.cuni.cz}

\usepackage[normalem]{ulem}

\def\ODdel{\bgroup\markoverwith{\textcolor{cyan!89!yellow!80!black!100}{\rule[0.4ex]{2pt}{3pt}}}\ULon}
\MakeRobust\ODdel

\begin{document}

\maketitle

\begin{abstract}
While recent neural sequence-to-sequence models have greatly improved the quality of speech synthesis, 
there has not been a system capable of 
fast training, fast inference and
high-quality audio synthesis at the same time. 
We propose a student-teacher network 
capable of high-quality faster-than-real-time spectrogram synthesis, with low requirements on computational resources and fast training time.
We show that self-attention layers are not necessary for generation of high quality audio. 
We utilize simple convolutional blocks with residual connections in both student and teacher networks and use only a single attention layer in the teacher model.
Coupled with a MelGAN vocoder, our model's voice quality was rated significantly higher than Tacotron~2.
Our model can be efficiently trained on a single GPU and can run in real time even on a 
CPU.
We provide both our source code and audio samples in our GitHub repository.\footnote{\url{https://github.com/janvainer/speedyspeech}\label{fn:github}}
\end{abstract}


\section{Introduction}

Recent neural text-to-speech (TTS) systems based on the sequence-to-sequence approach, such as Tacotron~2 \cite{Tacotron2}, brought considerable quality improvements, but require relatively large amounts of training data and computational resources to train and operate.
Several works attempt to reduce the computational burden in various ways \cite{DeepVoice3,EfficientTTS,WaveRNN,FastSpeech}, but there is still a tradeoff between fast training times, fast inference, and output quality.

In this paper, we address the training efficiency of TTS systems as well as the inference speed and hardware requirements while sustaining good quality of synthesized audio.
We propose a fully convolutional, non-sequential approach to speech synthesis 
consisting of a teacher and a student network, similarly to FastSpeech \cite{FastSpeech}. The teacher network is an autoregressive 
convolutional network 
\citep{DeepVoice3, EfficientTTS} which is used to extract correct alignments between phonemes and corresponding audio frames.
The student network is a non-autoregressive, fully convolutional network 
\cite{FastSpeech} 
which encodes input phonemes, predicts the duration (number of audio frames needed) for each one, then decodes a mel-scale spectrogram based on phoneme encodings and durations.
We combine our student network with a pretrained MelGAN vocoder \cite{MelGan} to achieve fast and high-quality spectrogram inversion.

Our model can be trained on the LJ~Speech data \cite{LJSpeech} in under 40 hours on a single 8GB GPU and generates high-quality audio samples faster than real-time on both GPU and CPU.

Our contributions are as follows: (1) We simplify the teacher-student architecture of FastSpeech \cite{FastSpeech} and provide a fast and stable training procedure. 
We use a simpler, smaller and faster-to-train convolutional teacher model with a single attention layer instead of Transformer \cite{Transformer} used in FastSpeech. 
(2) We show that self-attention layers \cite{Transformer} in the student network are not needed for
high-quality speech synthesis. 
(3) We describe a simple data augmentation technique that can be used early in the training to make the teacher network robust to sequential error propagation.
(4) We show that our model significantly outperforms strong baselines while keeping speedy training and inference.

\section{Related Work}

TTS systems such as Deep Voice 3 \cite{DeepVoice3} and DCTTS \cite{EfficientTTS} try to speed up training by utilizing convolutional networks inside an encoder-decoder architecture similar to Tacotron~2 \cite{Tacotron2}. The model trains fast, but requires sequential inference, which is relatively slow with convolutional networks.
WaveRNN \cite{WaveRNN} applies various hardware optimizations and model pruning to achieve sequential inference speedup. However, training is sequential and therefore slow.
To avoid sequential inference altogether, FastSpeech \cite{FastSpeech} adapts a Transformer-like architecture \cite{Transformer} along with the idea of fertilities. 
It can synthesize spectrogram frames quickly in parallel, but requires training of many attention layers, which can be difficult and time-consuming. 
Approaches such as Parallel WaveNet  \cite{ParallelWaveNet} and ClariNet \cite{ClariNet} provide fast inference, but require significant computational resources to train the teacher models.

\section{Our Model}
\label{sec:model}

Our model uses phonemes as input and logarithmically scaled mel spectrograms as output. 
We first discuss the teacher network used to align phonemes to spectrogram frames, then the student network which uses this alignment as additional supervision when training to synthesize spectrograms.

\subsection{Teacher network -- Duration extraction}
\label{sec:teacher}

\begin{figure}[t]
    \centering
    \includegraphics[width=0.95\columnwidth]{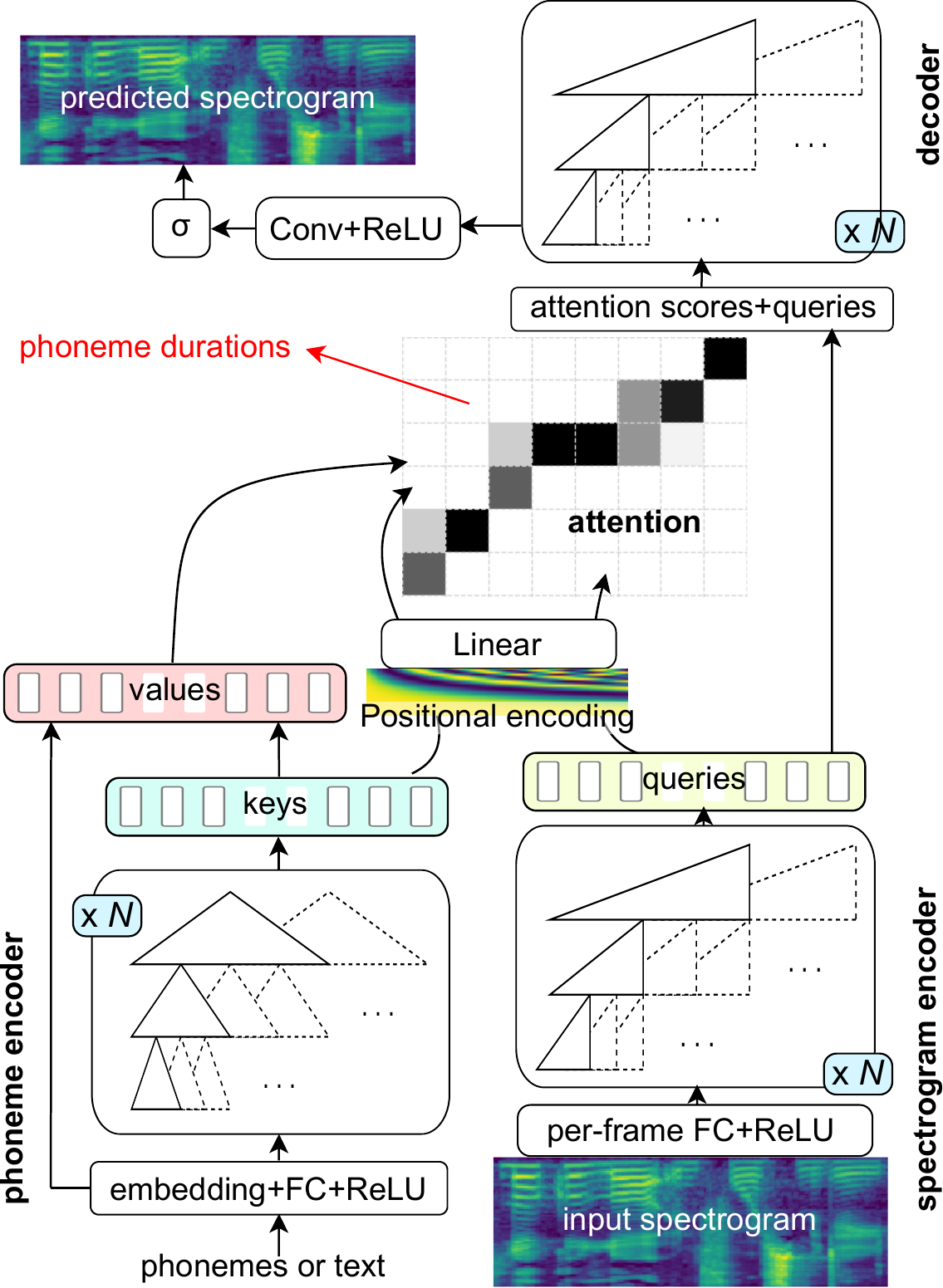}
    \caption{The duration extraction network. The encoders and the decoder use gated residual blocks (see Fig.~\ref{fig:wave_block}). Convolutions in the spectrogram encoder and decoder are causal 
    as the model predicts the next frame based on past ones
    (cf. Section~\ref{sec:teacher}).}
    \label{fig:duration-extract}
\end{figure}

The teacher network for extracting phoneme durations from data is 
based on Deep Voice~3 \cite{DeepVoice3} and DCTTS \cite{EfficientTTS}.
It has four main parts -- phoneme encoder, spectrogram encoder, attention and decoder (see Fig.~\ref{fig:duration-extract}).
It is trained to predict the next spectrogram frame given input phonemes (including punctuation) and past frames; it uses attention to keep track of the phoneme it is generating. The attention values are then used to align phonemes with spectrogram frames and extract phoneme durations.

\begin{figure}[t]
    \centering
    \includegraphics[width=0.7\columnwidth]{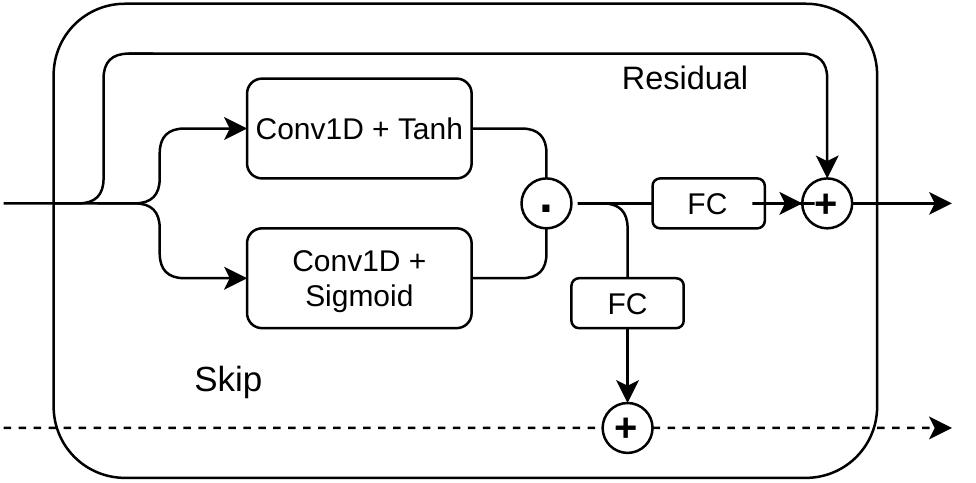}
    \caption{A gated residual block. 
    ``.'' and ``+'' represent element-wise multiplication and addition, respectively. 
    }
    \label{fig:wave_block}
\end{figure}

\paragraph*{Phoneme encoder: }The phoneme encoder
starts with embedding and a fully connected layer with ReLU activation.
Then, several gated residual blocks \cite[see Fig.~\ref{fig:wave_block}]{WaveNet} with progressively more dilated non-causal convolutions are used.
The blocks' skip connection sums outputs from all layers for the encoder output.

Instead of highway blocks used in DCTTS \citep{EfficientTTS}, we use these simple convolutional 
residual  blocks derived from WaveNet \citep{WaveNet} without observing any significant performance drop.

{}

\paragraph*{Spectrogram encoder: } The spectrogram encoder provides contextual encoding of spectrogram frames that takes past frames into account.
First, a fully connected layer and ReLU are applied to each frame of the input spectrogram.
Then, several gated residual blocks with progressively more dilated gated causal convolutions (over past frames only \cite{WaveNet}) are used and the skip connection accumulates the final output.

\paragraph*{Attention: }
We use dot-product attention \cite{Transformer}, with phoneme encoder output as keys, phoneme encoder outputs summed with phoneme embeddings as values (similar to Deep Voice 3 \cite{DeepVoice3}), and spectrogram encoder output as queries.
The keys and queries are preconditioned via positional encoding \cite{Transformer} and an identical linear layer to bias the attention towards monotonicity \cite[cf.~Section~\ref{sec:duration_extraction_experiments}]{DeepVoice3}. 
The attention scores are weighted averages of the value vectors according to how much the values match a given query. This way, the model learns to select phonemes relevant for prediction of the next spectrogram frame.

\paragraph*{Decoder: }
On the input, the decoder sums attention scores with encoder outputs for better gradient flow.
The sum is then
transformed by several gated residual blocks with progressively more dilated causal convolutions
and 
several convolutional layers with ReLU activation
to get the correct number of channels, and finally passed through a sigmoid prediction layer.

\paragraph*{Training:}
Target spectrograms are shifted one position to the left on the input and the model is forced to predict the next spectrogram frame based on input phonemes and previous frames. Unlike Tacotron~2 \cite{Tacotron2},  the network does not keep any hidden states and we can compute predictions for all time steps in parallel.
To be able to use the sigmoid activation in the final layer, we rescale the logarithmic mel spectrograms into the $[0,1]$ interval.

We minimize the sum of \emph{mean absolute error} (MAE) between the target and predicted spectrograms and \emph{guided attention loss} \cite{EfficientTTS}, which is used
to aid monotonic alignments. 
The guided attention loss for the attention matrix $A \in \mathbb{R}^{N \times T}$ is calculated as:
\begin{equation}
    GuidedAtt(A) = \frac{1}{NT}\sum_{n=1}^{N} \sum_{t=1}^{T}A_{n,t}W_{n,t}     
\end{equation}
where $W_{n,t} = 1 - \exp{-\frac{(n/N - t/T)^2}{2g^2}}$ is the penalty matrix, $N$ is the number of phonemes and $T$ is the number of spectrogram frames.
The parameter $g$ controls the loss contribution of matrix elements $A_{n,t}$ as we move further away from the diagonal.

\paragraph*{Data augmentation:}
To improve robustness to error propagation, we employ three data augmentations on the input spectrograms: 
(1) We add a small amount of Gaussian noise to each spectrogram pixel.
(2) We simulate the model outputs by feeding the input spectrogram through the network without gradient update in parallel mode (not sequentially). The resulting spectrogram is slightly degraded compared to the ground-truth spectrogram. 
We repeat this process multiple times to get an approximation of a sequentially generated spectrogram.
We could simply generate the degraded spectrogram sequentially, but using the parallel mode several times is still faster than sequential generation. Moreover, in early stages of training, the model is virtually unable to sequentially generate more than just a few frames correctly. 
We observe that this method improves the robustness of sequential generation drastically and the model is able to generate long sentences well with just minor mistakes.
(3) We further degrade the input spectrograms by randomly replacing several frames with random frames. This is done to encourage the model to use temporally more distant frames. Otherwise, the model tends to overfit to the newest frame on the input and ignores older information, which makes it less stable.


\paragraph*{Inference/duration extraction:}
Similarly to \cite{DeepVoice3, EfficientTTS}, we apply location masking of the attention positions to avoid phoneme skipping and enforce monotonic alignment. However, we run the inference in teacher-forcing mode -- 
we feed the model with ground-truth frames to avoid error propagation and extract more reliable alignments.
The resulting attention matrix is used to extract the duration of each phoneme by calculating the index of the most likely phoneme at each timestep and counting the number of occurrences of each index across time.

\subsection{Student network -- Spectrogram synthesis}
\label{sec:student}

\begin{figure}[t]
    \centering
    \includegraphics[width=\columnwidth]{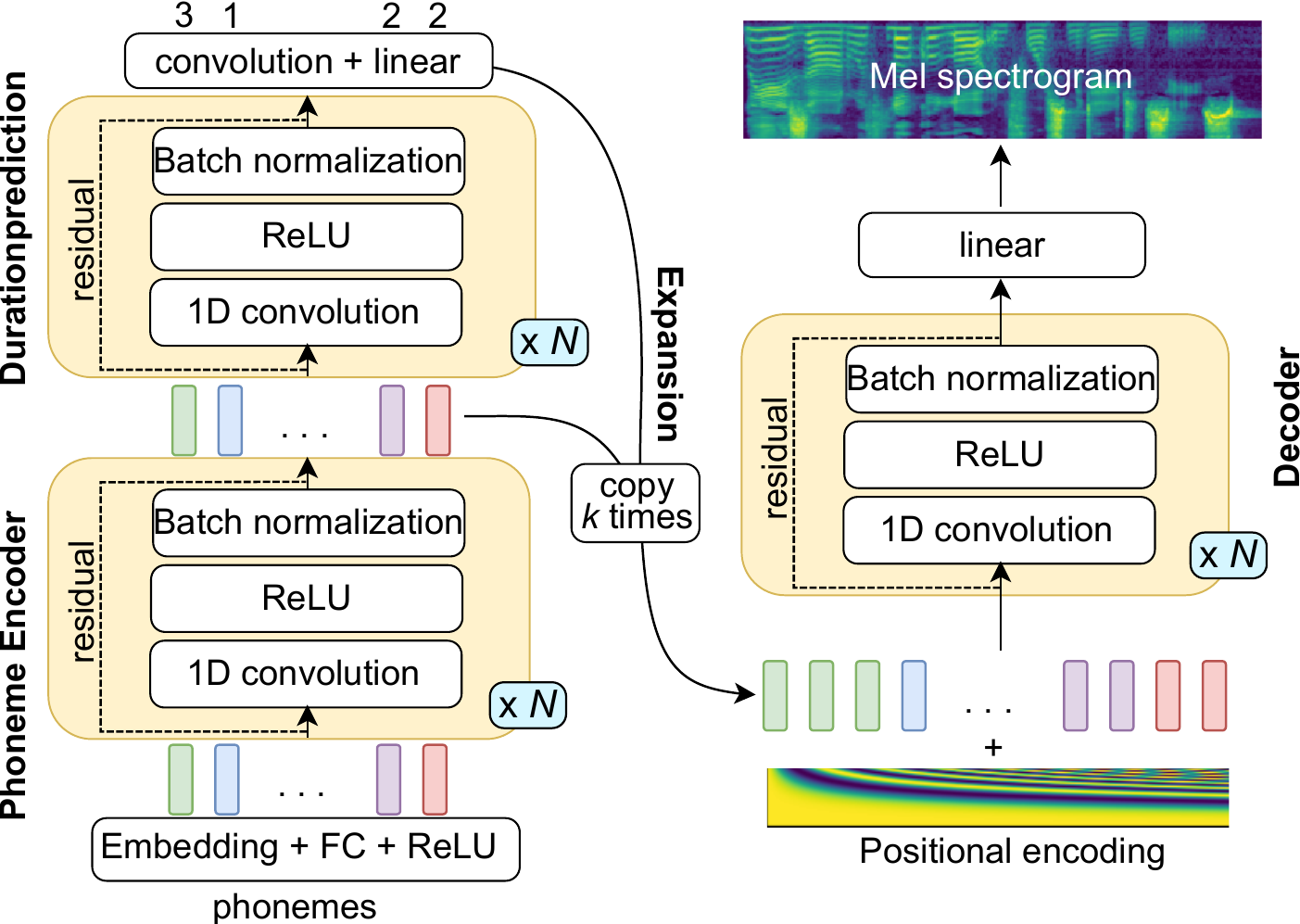}
    \caption{The spectrogram synthesis model. 
    The duration predictor predicts a phoneme's duration (number of frames) based on its encoding. 
    The encoding is copied for the predicted number of times on the decoder input (cf.~Section~\ref{sec:student}).
    }
    \label{fig:speedy_speech}
\end{figure}


The student model uses spectrograms with alignments predicted by the teacher model. Given input phonemes, it is trained to first predict individual phoneme durations and then, based on the durations, the full mel spectrogram (see Fig.~\ref{fig:speedy_speech}).
The model consists of a phoneme encoder, duration predictor and a decoder. All three modules consist of progressively dilated  residual convolutional blocks, each of which contains a 1D convolution, ReLU activation and temporal batch normalization. A residual connection is applied for better gradient flow. The phoneme encodings generated by the encoder are fed to the duration predictor, which predicts the duration of each phoneme in a logarithmic domain using a final convolution and a linear layer.  

Phoneme encoding vectors are expanded according to the predicted duration on the decoder input so that the size of the decoder input matches the desired size of the output spectrogram.
Similarly to FastSpeech \cite{FastSpeech}, we add positional encoding \citep{Transformer} to the phoneme encoding vectors, but we reset the encoding for each phoneme. We hypothesize that it is more beneficial for the network to distinguish the frame location in the context of a single phoneme instead of the whole sentence. 
The decoder converts the expanded phoneme encodings with positional embeddings into individual frames of a mel spectrogram.

Our student model is inspired by FastSpeech \cite{FastSpeech},
but we replace attention with residual convolutional blocks and use temporal batch normalization instead of layer normalization.

\paragraph*{Training:}
We use the sum of MAE and \emph{structural similarity index} (SSIM) \citep{Wang2004} losses for logarithmic mel spectrogram value regression and \emph{Huber loss} \cite[p.~349]{hastie_elements_2009} for logarithmic duration prediction.
%
%
We use the ground-truth durations extracted with the teacher model during training for the phoneme encoding expansion. We found it beneficial to normalize the target logarithmic mel spectrograms to have zero mean and unit variance.
Unlike FastSpeech~\cite{FastSpeech}, we detach gradient flow from the duration predictor to the encoder; this increased performance of spectrogram prediction and reduced overfitting of the duration predictor.


\section{Experimental Setup}
\label{sec:experiments}

Here we describe 
our dataset and training process,
including our preliminary experiments that led to selecting model parameters.

\subsection{Dataset}

We train our model on the publicly available LJ Speech dataset \citep{LJSpeech}, which consists of 13,100 recordings
and corresponding transcripts of a single professional female speaker reading from several English texts. 
Numbers 
and monetary units in the transcription are expanded into full words. 
We reserve the last 100 utterances for evaluation, the rest is used for training.

We phonemize the transcripts with the g2p python package,\footnote{\url{https://github.com/Kyubyong/g2p}} and use phonemic transcription as the input to both our teacher and student network.
We transform linear spectrograms to mel scale and a log transformation is applied on the amplitudes.

\subsection{Teacher network parameters}
\label{sec:duration_extraction_experiments}

We settled on using 10 residual blocks for both encoders and 14 residual blocks for the decoder. 
We used kernel size 3 and dilation rates 1, 3, 9, 27, 1, 3, 9, 27 for the first 8 blocks, with dilation 1 for the remaining ones.
We used 40 channels for the skip connections and 80 channels for the gates.

We used the Adam optimizer \cite{kingma_adam:_2015} with default parameters and gradient clipping at 1. We tried different learning rates and schedules and settled on 0.002 base rate with inverse square root decay with a 30-epoch warmup (Noam scheduler) \cite{Transformer}, which provided the best tradeoff in terms of stability and speed. 



We found 
that attention learning is considerably faster when guided attention loss \cite{EfficientTTS} and especially attention preconditioning with positional encoding \cite{DeepVoice3} are applied -- both measures aim at near monotonic attention.
Plain attention takes around 100 epochs of training to become near-monotonic; with the guided attention loss, this comes down to 50 epochs. Attention preconditioning assumes monotonic attention from the very beginning.
We further tried to improve attention robustness by applying dropout \cite{Dropout}, but this did not bring any improvements.



\subsection{Student network parameters}
\label{sec:spectrogram_synthesis_experiments}

The student model is generally very sensitive to the teacher network's duration extraction accuracy; without accurate phoneme durations, it will not converge. 
We experimented with inverse square root decay and reduce-on-plateau \cite{zaheer_adaptive_2018} schedules and settled for the latter and a base learning rate 0.002. Adam optimizer with default parameters and gradient clipping was used. 

We observed that the network depth and dilation factor must be high enough to span more than a single phoneme. This is caused by segments of identical vectors on the decoder input -- phoneme encodings are copied multiple times
to compensate for the length mismatch of the input phoneme sequence and the output spectrogram.
Applying a short convolution on a sequence consisting of homogeneous segments will then result in another sequence of largely homogeneous segments. 
Therefore, we used 26 encoder blocks with dilations repeating the pattern 1, 1, 2, 2, 4, 4, three duration predictor blocks with dilations 4, 3, 1 and 34 decoder blocks with dilations repeating the pattern 1, 1, 2, 2, 4, 4, 8, 8.
We used 128 channels in all convolutional layers.

We use batch normalization \citep{BatchNorm}, i.e., per-channel normalization across all time steps and items in a batch, as we found it to alleviate the vanishing gradient problem and speed up training.
We also tried layer normalization \citep{LayerNorm}, channel normalization without considering the batch dimension, or dropout applied after normalization, but none of this brought any further benefits. 

We compared our decoder to a variant trained without using the SSIM loss. 
This produced blurrier spectrograms, 
but the difference in audio quality was not noticeable. 
Higher-quality vocoding might make this issue visible. 
 



We compared our local position encodings in the decoder to global position encodings and no position encoding.
We found that local encodings only bring very small benefits in terms of $L_1$ and SSIM loss, but still decided to use them in the final model.



\section{Evaluation}
\label{sec:results}

We evaluate our model in terms of subjective voice quality perception, inference speed and time required for training.

\subsection{Voice quality}

\begin{table}[t]
    \caption{Resulting MUSHRA-like scores from our survey, with 95\% confidence intervals calculated with bootstrap resampling.}
    \label{table:eval_quality}
    \centering
    \begin{tabular}{lcc}
    \hline
        \bf Model (vocoding)               & \bf Mean Score  & \bf 95 \% CI      \\
        \hline
        Tacotron~2 (MelGAN)         & 62.82 & $(-2.01, +2.20)$  \\
        Deep Voice~3 (lws)         & 43.61 & $(-2.25, +2.20)$\\
        Reference           & 97.85 & $(-0.76, +0.66)$  \\
        \hline
        Ours (Griffin-Lim) & 47.03 & $(-2.00, +2.16)$ \\
        Ours (MelGAN)           & $\mathbf{75.24}$  & $(-1.91, +1.73)$ \\\hline

    \end{tabular}
\end{table}{}

To measure and compare quality of the synthesized audio, we conducted an in-house survey with 40 participants. 
We synthesized our LJ Speech held-out sentences 
with our model 
and several baselines trained on the same data.
The 
ground-truth recordings were used as a reference.

We used a setting based on MUSHRA~\cite{ITU-R,WebMushra}: 
the participants were shown anonymized outputs of all models and the reference for a given sentence, and they rated them on a fine-grained 100-point scale,
visually divided into 5 categories: “Excellent”, “Fair”, “Good”, “Poor” and “Bad”. 
Unlike MUSHRA, we did not use anchor recordings.
We discarded any participants who rated the reference 
under 90 in 8 or more 
cases out of 10.

We selected audio examples produced by the following setups for comparison:\footnote{We do not compare against FastSpeech \cite{FastSpeech} as no implementation of this model was available to us.}
\begin{itemize}[nosep]
    \item Reference human audio recording
    \item Deep Voice 3 \cite{DeepVoice3}\footnote{Implementation used: \url{https://github.com/r9y9/deepvoice3_pytorch}} + lws \cite{LWS} vocoding
    \item Tacotron~2 \cite{Tacotron2}\footnote{Implementation used: \url{https://github.com/NVIDIA/tacotron2}} + MelGAN\footnote{Code + checkpoint used: \url{https://github.com/seungwonpark/melgan}} vocoding
    \item Ours + Griffin-Lim \cite{Griffin} vocoding
    \item Ours + MelGAN vocoding
\end{itemize}{}
We offer two versions of our model for a fair comparison with the baselines' vocoders.\footnote{Due to incompatibility of STFT implementations, 
we were not able to use lws for vocoding 
with our model. However, we provide a version that uses Griffin-Lim, a weaker-performing signal estimation algorithm.}

The results 
are displayed in Table~\ref{table:eval_quality}. 
We used bootstrap resampling \citep{BootstrapResampling} to obtain the mean and 95\% confidence intervals.\footnote{The resampling was done 1000 times.}
Our model with MelGAN attained the average score of 75 
and scored significantly higher than Tacotron~2. 
Our model with Griffin-Lim was also able to achieve a significantly higher score than Deep Voice 3 with lws.
This shows that our model is clearly preferred to both baselines when used with a similar vocoder.

On manual analysis of the outputs, we found fewer pronunciation mistakes and better intonation consistency in our model compared to the baselines.
We account this to the fact that the baseline models are both sequential
and condition on past spectrogram frames, but do not have access to future ones.
This can make the spectrograms more locally accurate, but the global consistency may be lower.
In contrast, our model does not condition generation on past frames as all frames are generated in parallel, but is able to aggregate information across the entire input.

\subsection{Inference speed}

We measured inference speed for different batch sizes, created by repeating the same input (34 words, 112 phonemes, 9.72 seconds on the output, see Table~\ref{table:eval_inference_speed}).

\begin{table}[t]
    \caption{Inference time for batches of different size on a 4GB GeForce GTX 960M GPU (left) and Intel Core i5-6300HQ 2.3 GHz 4-core CPU (right), averaged over 10 runs: times in seconds to produce the spectrogram, the waveform (audio) and the total. 
    Each produced sample in the batch is $9.72$ seconds long.} 
    \label{table:eval_inference_speed}
    \centering
    \newcommand{\pz}{\phantom{0}} 
    \begin{tabular}{>{\hspace{-2mm}}cc>{\hspace{-2mm}}c>{\hspace{-2mm}}cc>{\hspace{-2mm}}c>{\hspace{-2mm}}c<{\hspace{-2mm}}}
        \hline
        \bf Batch & \multicolumn{3}{c}{\bf GPU} & \multicolumn{3}{c}{\bf CPU} \\
        \bf size & \bf S-gram & \bf Audio & \bf Total & \bf S-gram & \bf Audio & \bf Total \\
        \hline
        1  & 0.032 & 0.165 & 0.197 & 0.105 & \pz1.702  & \pz1.808  \\
        2  & 0.035 & 0.325 & 0.359 & 0.137 & \pz3.211  & \pz3.348  \\
        4  & 0.050 & 0.647 & 0.697 & 0.263 & \pz6.788  & \pz7.051  \\
        8  & 0.097 & 1.291 & 1.388 & 0.591 & 14.061 & 14.652 \\
        16 & 0.203 & 4.065 & 4.268 & 1.219 & 27.685 & 28.904 \\
        \hline 
    \end{tabular}{}
\end{table}{}

We are able to synthesize $9.72$s of audio in 197ms on a GPU, which is 49$\times$ faster than real time (and about 8.8$\times$ faster overall than Tacotron 2 on the same GPU, with the spectrogram generation step being 48.5$\times$ faster). 
On a CPU, we are able to synthesize approximately $5 \times$ faster than real time. 
Synthesizing batches, we are able to synthesize $16 \times 9.72 = 155.52$s of audio in $4.27$s on a GPU, which is over 36 times faster than real time.
Our model 
scales well even on a CPU without advanced optimization such as weight pruning or weight quantization.

\subsection{Training time}

Both the duration extraction (teacher) and spectrogram synthesis (student) models 
were trained on a single GeForce GTX 1080 GPU with 8GB RAM, with batch size 64. 
The training times along with the total number of model parameters are shown in Table~\ref{table:eval_training_speed}.
The teacher model is smaller, but takes longer to train since a smaller learning rate must be used to converge with good results (see Section~\ref{sec:experiments}).
The student model is larger, but the architecture is simpler and does not contain any hard-to-train components such as attention, which makes it converge easier.

\begin{table}[t]
    \caption{Model size and training speed 
    for the duration extraction (teacher) and the spectrogram synthesis (student) models, measured on a single GeForce GTX 1080 GPU with 8GB RAM. 
    }
    \label{table:eval_training_speed}
    \centering
    \begin{tabular}{lcc}
        \hline
            & \bf Teacher & \bf Student \\
        \hline 
         Total parameters & 708,920 & 4,306,001 \\
         Training time (hours) & 19 & 13 \\
         Epochs till convergence & 250 & 100 \\
         Time per epoch (minutes)  & 4.56 & 7.8 \\
        \hline 
    \end{tabular}{}
\end{table}{}

\section{Conclusion}

We presented a convolutional model for spectrogram synthesis from phonemes that supports both speedy training and inference, while maintaining significantly better output voice quality than strong baselines.
Our source code and audio samples are available on GitHub.\textsuperscript{\ref{fn:github}}
For future work, we plan to extend the model to support multi-speaker training data.

\section{Acknowledgements}

This research was supported by the Charles University grant PRIMUS/19/SCI/10.

\renewcommand\bibsection{\section{References}}
\bibliographystyle{IEEEtranN}
\bibliography{mybib}

\begin{thebibliography}{23}
\providecommand{\natexlab}[1]{#1}
\providecommand{\url}[1]{#1}
\csname url@samestyle\endcsname
\providecommand{\newblock}{\relax}
\providecommand{\bibinfo}[2]{#2}
\providecommand{\BIBentrySTDinterwordspacing}{\spaceskip=0pt\relax}
\providecommand{\BIBentryALTinterwordstretchfactor}{4}
\providecommand{\BIBentryALTinterwordspacing}{\spaceskip=\fontdimen2\font plus
\BIBentryALTinterwordstretchfactor\fontdimen3\font minus
  \fontdimen4\font\relax}
\providecommand{\BIBforeignlanguage}[2]{{%
\expandafter\ifx\csname l@#1\endcsname\relax
\typeout{** WARNING: IEEEtranN.bst: No hyphenation pattern has been}%
\typeout{** loaded for the language `#1'. Using the pattern for}%
\typeout{** the default language instead.}%
\else
\language=\csname l@#1\endcsname
\fi
#2}}
\providecommand{\BIBdecl}{\relax}
\BIBdecl

\bibitem[Shen et~al.(2018)Shen, Pang, Weiss, Schuster, Jaitly, Yang, Chen,
  Zhang, Wang, Skerrv-Ryan, Saurous, Agiomvrgiannakis, and Wu]{Tacotron2}
J.~Shen, R.~Pang, R.~J. Weiss, M.~Schuster, N.~Jaitly, Z.~Yang, Z.~Chen,
  Y.~Zhang, Y.~Wang, R.~Skerrv-Ryan, R.~A. Saurous, Y.~Agiomvrgiannakis, and
  Y.~Wu, ``{Natural TTS Synthesis by Conditioning Wavenet on MEL Spectrogram
  Predictions},'' in \emph{Proceedings of the IEEE International Conference on
  Acoustics, Speech and Signal Processing (ICASSP)}, Calgary, AB, Canada, Apr.
  2018, pp. 4779--4783.

\bibitem[Ping et~al.(2018)Ping, Peng, Gibiansky, Arık, Kannan, Narang, Raiman,
  and Miller]{DeepVoice3}
W.~Ping, K.~Peng, A.~Gibiansky, S.~Arık, A.~Kannan, S.~Narang, J.~Raiman, and
  J.~Miller, ``{Deep Voice 3: Scaling text-to-speech with convolutional
  sequence learning},'' in \emph{Proceedings of the 6th International
  Conference on Learning Representations (ICLR)}, Vancouver, BC, Canada, Oct.
  2018.

\bibitem[Tachibana et~al.(2018)Tachibana, Uenoyama, and Aihara]{EfficientTTS}
H.~Tachibana, K.~Uenoyama, and S.~Aihara, ``{Efficiently Trainable
  Text-to-Speech System Based on Deep Convolutional Networks with Guided
  Attention},'' in \emph{Proceedings of the IEEE International Conference on
  Acoustics, Speech and Signal Processing - Proceedings (ICASSP)}, Calgary, AB,
  Canada, Apr. 2018, pp. 4784--4788.

\bibitem[Kalchbrenner et~al.(2018)Kalchbrenner, Elsen, Simonyan, Noury,
  Casagrande, Lockhart, Stimberg, Oord, Dieleman, and Kavukcuoglu]{WaveRNN}
N.~Kalchbrenner, E.~Elsen, K.~Simonyan, S.~Noury, N.~Casagrande, E.~Lockhart,
  F.~Stimberg, A.~v.~d. Oord, S.~Dieleman, and K.~Kavukcuoglu, ``Efficient
  {Neural} {Audio} {Synthesis},'' in \emph{Proceedings of the 35th
  {International} {Conference} on {Machine} {Learning} (ICML)}, Stockholm,
  Sweden, Jul. 2018, pp. 2410--2419.

\bibitem[Ren et~al.(2019)Ren, Ruan, Tan, Qin, Zhao, Zhao, and Liu]{FastSpeech}
Y.~Ren, Y.~Ruan, X.~Tan, T.~Qin, S.~Zhao, Z.~Zhao, and T.-Y. Liu,
  ``{FastSpeech: Fast, Robust and Controllable Text to Speech},'' in
  \emph{Advances in Neural Information Processing Systems 32 (NeurIPS)},
  Vancouver, BC, Canada, Dec. 2019, pp. 3171--3180.

\bibitem[Kumar et~al.(2019)Kumar, Kumar, de~Boissiere, Gestin, Teoh, Sotelo,
  de~Brebisson, Bengio, and Courville]{MelGan}
K.~Kumar, R.~Kumar, T.~de~Boissiere, L.~Gestin, W.~Z. Teoh, J.~Sotelo,
  A.~de~Brebisson, Y.~Bengio, and A.~Courville, ``{MelGAN: Generative
  Adversarial Networks for Conditional Waveform Synthesis},'' in \emph{Advances
  in Neural Information Processing Systems 32 (NeurIPS)}, Vancouver, BC,
  Canada, Dec. 2019, pp. 14\,910--14\,921.

\bibitem[Ito(2017)]{LJSpeech}
\BIBentryALTinterwordspacing
K.~Ito, ``{The LJ Speech Dataset},'' 2017. [Online]. Available:
  \url{https://keithito.com/LJ-Speech-Dataset/}
\BIBentrySTDinterwordspacing

\bibitem[Vaswani et~al.(2017)Vaswani, Shazeer, Parmar, Uszkoreit, Jones, Gomez,
  Kaiser, and Polosukhin]{Transformer}
A.~Vaswani, N.~Shazeer, N.~Parmar, J.~Uszkoreit, L.~Jones, A.~N. Gomez,
  {\L}.~Kaiser, and I.~Polosukhin, ``{Attention is all you need},'' in
  \emph{Advances in Neural Information Processing Systems 30 (NeurIPS)}, Long
  Beach, CA, USA, Dec. 2017, pp. 5999--6009.

\bibitem[Van Den~Oord et~al.(2018)Van Den~Oord, Li, Babuschkin, Simonyan,
  Vinyals, Kavukcuoglu, Van Den~Driessche, Lockhart, Cobo, Stimberg,
  Casagrande, Grewe, Noury, Dieleman, Elsen, Kalchbrenner, Zen, Graves, King,
  Walters, Belov, and Hassabis]{ParallelWaveNet}
A.~Van Den~Oord, Y.~Li, I.~Babuschkin, K.~Simonyan, O.~Vinyals, K.~Kavukcuoglu,
  G.~Van Den~Driessche, E.~Lockhart, L.~C. Cobo, F.~Stimberg, N.~Casagrande,
  D.~Grewe, S.~Noury, S.~Dieleman, E.~Elsen, N.~Kalchbrenner, H.~Zen,
  A.~Graves, H.~King, T.~Walters, D.~Belov, and D.~Hassabis, ``{Parallel
  WaveNet: Fast high-fidelity speech synthesis},'' in \emph{Proceedings of the
  35th International Conference on Machine Learning (ICML)}, Stockholm, Sweden,
  Jul. 2018, pp. 6270--6278.

\bibitem[Ping et~al.(2019)Ping, Peng, and Chen]{ClariNet}
W.~Ping, K.~Peng, and J.~Chen, ``Clarinet: Parallel wave generation in
  end-to-end text-to-speech,'' in \emph{Proceedings of the Seventh
  International Conference on Learning Representations (ICLR)}, New Orleans,
  LA, USA, May 2019.

\bibitem[Oord et~al.(2016)Oord, Dieleman, Zen, Simonyan, Vinyals, Graves,
  Kalchbrenner, Senior, and Kavukcuoglu]{WaveNet}
A.~v.~d. Oord, S.~Dieleman, H.~Zen, K.~Simonyan, O.~Vinyals, A.~Graves,
  N.~Kalchbrenner, A.~Senior, and K.~Kavukcuoglu, ``{WaveNet: A Generative
  Model for Raw Audio},'' \emph{arXiv preprint arXiv:1609.03499}, Sep. 2016.

\bibitem[Wang et~al.(2004)Wang, Bovik, Sheikh, and Simoncelli]{Wang2004}
Z.~Wang, A.~C. Bovik, H.~R. Sheikh, and E.~P. Simoncelli, ``{Image quality
  assessment: From error visibility to structural similarity},'' \emph{IEEE
  Transactions on Image Processing}, vol.~13, no.~4, pp. 600--612, Apr. 2004.

\bibitem[Hastie et~al.(2009)Hastie, Tibshirani, Friedman, and
  Franklin]{hastie_elements_2009}
T.~Hastie, R.~Tibshirani, J.~Friedman, and J.~Franklin, \emph{The elements of
  statistical learning: data mining, inference and prediction}, 2nd~ed.\hskip
  1em plus 0.5em minus 0.4em\relax Springer, 2009.

\bibitem[Kingma and Ba(2015)]{kingma_adam:_2015}
D.~Kingma and J.~Ba, ``Adam: {A} {Method} for {Stochastic} {Optimization},'' in
  \emph{Proceedings of the 3rd {International} {Conference} on {Learning}
  {Representations} (ICLR)}, San Diego, CA, USA, May 2015.

\bibitem[Srivastava et~al.(2014)Srivastava, Hinton, Krizhevsky, Sutskever, and
  Salakhutdinov]{Dropout}
N.~Srivastava, G.~Hinton, A.~Krizhevsky, I.~Sutskever, and R.~Salakhutdinov,
  ``Dropout: A simple way to prevent neural networks from overfitting,''
  \emph{Journal of Machine Learning Research}, vol.~15, no.~56, pp. 1929--1958,
  Jun. 2014.

\bibitem[Zaheer et~al.(2018)Zaheer, Reddi, Sachan, Kale, and
  Kumar]{zaheer_adaptive_2018}
M.~Zaheer, S.~Reddi, D.~Sachan, S.~Kale, and S.~Kumar, ``Adaptive {Methods} for
  {Nonconvex} {Optimization},'' in \emph{Advances in {Neural} {Information}
  {Processing} {Systems} 31 (NeurIPS)}, S.~Bengio, H.~Wallach, H.~Larochelle,
  K.~Grauman, N.~Cesa-Bianchi, and R.~Garnett, Eds., Montréal, QC, Canada,
  Dec. 2018, pp. 9793--9803.

\bibitem[Ioffe and Szegedy(2015)]{BatchNorm}
S.~Ioffe and C.~Szegedy, ``Batch normalization: accelerating deep network
  training by reducing internal covariate shift,'' in \emph{Proceedings of the
  32nd International Conference on Machine Learning (ICML)}, Lille, France,
  Jul. 2015, pp. 448--456.

\bibitem[Ba et~al.(2016)Ba, Kiros, and Hinton]{LayerNorm}
J.~L. Ba, J.~R. Kiros, and G.~E. Hinton, ``Layer normalization,'' in
  \emph{Proceedings of the Neural Information Processing Systems Deep Learning
  Symposium}, Barcelona, Spain, Dec. 2016.

\bibitem[ITU(2015)]{ITU-R}
``{Method for the subjective assessment of intermediate quality level of audio
  systems},'' {International Telecommunication Union}, Geneva, Recommendation
  BS.1534, 2015.

\bibitem[Schoeffler et~al.(2018)Schoeffler, Bartoschek, St{\"{o}}ter, Roess,
  Westphal, Edler, and Herre]{WebMushra}
M.~Schoeffler, S.~Bartoschek, F.-R. St{\"{o}}ter, M.~Roess, S.~Westphal,
  B.~Edler, and J.~Herre, ``{webMUSHRA — A Comprehensive Framework for
  Web-based Listening Tests},'' \emph{Journal of Open Research Software},
  vol.~6, no.~1, p.~8, Feb. 2018.

\bibitem[Le~Roux et~al.(2010)Le~Roux, Kameoka, Ono, and Sagayama]{LWS}
J.~Le~Roux, H.~Kameoka, N.~Ono, and S.~Sagayama, ``{Fast signal reconstruction
  from magnitude STFT spectrogram based on spectrogram consistency},'' in
  \emph{Proceeedings of the 13th International Conference on Digital Audio
  Effects (DAFx)}, Graz, Austria, Sep. 2010.

\bibitem[Griffin and Lim(1984)]{Griffin}
D.~W. Griffin and J.~S. Lim, ``{Signal Estimation from Modified Short-Time
  Fourier Transform},'' \emph{IEEE Transactions on Acoustics, Speech, and
  Signal Processing}, vol.~32, no.~2, pp. 236--243, Apr. 1984.

\bibitem[Efron(1979)]{BootstrapResampling}
B.~Efron, ``{Bootstrap Methods: Another Look at the Jackknife},'' \emph{The
  Annals of Statistics}, vol.~7, no.~1, pp. 1--26, Jan. 1979.

\end{thebibliography}

\end{document}